\documentclass[journal]{IEEEtran}
\ifCLASSINFOpdf
\else
\fi

\usepackage{graphicx}
\usepackage{bm}
\usepackage{amsfonts}
\usepackage{amsmath}
\usepackage{amssymb}
\usepackage{times}
\usepackage{subfigure}
\usepackage{latexsym,bm,amsmath,amssymb} 
\usepackage{cite}
\usepackage{hhline}

\usepackage{multirow} 
\usepackage{amsmath}
\usepackage{xcolor}
\usepackage{epstopdf}
\usepackage{algorithmicx,algorithm}

\usepackage{geometry}
\begin{document}

%
\title{Orthogonal Time Frequency Space Modulation -- Part II: Transceiver Designs}

\author{Shuangyang~Li,~\IEEEmembership{Member,~IEEE,}
Weijie~Yuan,~\IEEEmembership{Member,~IEEE,}
Zhiqiang~Wei,~\IEEEmembership{Member,~IEEE,}
Robert~Schober,~\IEEEmembership{Fellow,~IEEE,}
and
Giuseppe~Caire,~\IEEEmembership{Fellow,~IEEE}\\
{(\emph{Invited Paper})}

\thanks{S.~Li was with the School of Electrical Engineering and Telecommunications, University of New South Wales, Sydney, NSW
2052, Australia, when this letter was submitted. He is now with the Department of Electrical, Electronic, and Computer Engineering, University of Western Australia, Perth, WA 6009, Australia (e-mail: shuangyang.li@uwa.edu.au).}
\thanks{
W.~Yuan is with the Department of Electrical and Electronic Engineering,
Southern University of Science and Technology, Shenzhen 518055, China
(e-mail: yuanwj@sustech.edu.cn).}
\thanks{Z. Wei is with the School of Mathematics and Statistics, Xi'an Jiaotong University, Xi'an 710049, China (e-mail: zhiqiang.wei@xjtu.edu.au).}
\thanks{R.~Schober is with the Institute for Digital Communications (IDC), the Friedrich-Alexander University Erlangen-Nuremberg, Erlangen 91054, Germany (e-mail: {robert.schober}@fau.de).}
\thanks{G. Caire is with the Electrical Engineering and Computer Science Department, the Technische Universit{\"a}t Berlin, Berlin 10587, Germany (e-mail: caire@tu-berlin.de).}
\vspace{-8mm}
}

\maketitle

\begin{abstract}
The fundamental concepts and challenges of orthogonal time frequency space (OTFS) modulation have been reviewed in Part I of this three-part tutorial. In this second part, we provide an overview of the state-of-the-art transceiver designs for OTFS systems, with a particular focus on the cyclic prefix (CP) design, window design, pulse shaping, channel estimation, and signal detection. Furthermore, we analyze the performance of OTFS modulation, including the diversity gain and the achievable rate. Specifically,
comparative simulations are presented to evaluate the error performance of different OTFS detection schemes, and the advantages of coded OTFS systems over coded orthogonal frequency-division multiplexing (OFDM) systems are investigated.
\end{abstract}


\IEEEpeerreviewmaketitle

\vspace{-8mm}
\section{Introduction}
Orthogonal time frequency space (OTFS) modulation has received considerable attention in the past few years since its introduction in~\cite{Hadani2017orthogonal}, thanks to its capability of enabling highly reliable communication over high-mobility channels~\cite{Zhiqiang_magzine}.
The most important new feature of OTFS modulation compared to conventional orthogonal frequency-division multiplexing (OFDM) modulation is the delay-Doppler (DD) domain information multiplexing, which motivates OTFS transceiver design based on the DD domain channel response. Consequently, conventional transceiver designs for OFDM systems optimized based on the time-frequency (TF) domain channel characteristics cannot be directly applied in OTFS systems as they are not able to harvest the full benefits of DD domain information multiplexing.

In Part II of this three-part tutorial, we aim to provide an in-depth discussion on OTFS transceiver design. Specifically, we study the key elements of the transceiver, including cyclic prefix (CP) insertion, pulse shaping, channel estimation, and signal detection. In particular, the commonly used message passing algorithm (MPA) for OTFS detection is explained based on the \emph{maximum a posteriori} (MAP) criterion, and simulation results are presented to evaluate the error performance of various detection schemes. 
Furthermore, we compare the performances of OTFS and OFDM in terms of diversity gain and achievable rate, where we also numerically verify the advantages of coded OTFS modulation over coded OFDM.

\section{Transmitter Design}
The transmitter design is of great importance for practical application of OTFS. As explained in Part I, there are two common implementations of OTFS, namely, symplectic finite Fourier transform (SFFT)-based OTFS and discrete Zak transform (DZT)-based OTFS. In this section, we will provide further details on the transmitter design for both SFFT-based and DZT-based OTFS, respectively.

Similar to the Part I, we assume that one OTFS frame occupies a bandwidth of $B_{\rm OTFS}$ and a time duration of $T_{\rm OTFS}$, which accommodates $M$ subcarriers with subcarrier spacing $\Delta f = \frac{{{B_{{\rm{OTFS}}}}}}{M}$ and $N$ time slots with slot duration $T = \frac{{{T_{{\rm{OTFS}}}}}}{N}$.
\subsection{Cyclic Prefix Design for SFFT-based OTFS}
The SFFT-based implementation was proposed in the first OTFS paper~\cite{Hadani2017orthogonal}. In particular, the SFFT-based implementation can be viewed as the concatenation of an inverse SFFT (ISFFT) module and the Heisenberg transform, where the latter one can be realized with an inverse fast Fourier transform (IFFT) module followed by a transmit pulse shaping filter~\cite{Hadani2017orthogonal}.

The details of SFFT-based OTFS have been covered in Part~I. Here,
we focus on CP design. Specifically, there are two commonly used options for inserting the CP into SFFT-based OTFS, i.e., \textit{full-CP} OTFS and \textit{reduced-CP} OTFS.
In the full-CP scheme, a CP is inserted in each time slot to combat the delay spread of the channel, similar to what is done in conventional OFDM~\cite{RezazadehReyhani2018analysis}. On the other hand,
in the reduced-CP scheme, only one CP is appended at the start of the frame
with a duration longer than the maximum delay spread of the channel. Reduced-CP OTFS has been officially introduced in the literature in~\cite{Raviteja2019practical}.

A key property of full-CP OTFS is that intersymbol interference (ISI)-free transmission can be guaranteed after CP removal at the receiver side, similar to conventional OFDM~\cite{RezazadehReyhani2018analysis}. As a result,
signal detection can be performed in the TF domain, where only the impact of the Doppler shifts of the channel has to be considered. Therefore, full-CP OTFS transmissions may enable reduced-complexity signal detection.


On the other hand, reduced-CP OTFS may be the more attractive option. In contrast to full-CP OTFS, the reduced-CP scheme does not guarantee ISI-free transmission, but it generally requires a much smaller signaling overhead.
In fact, the purpose of the reduced-CP scheme is to ensure that the received sequence is $MN$-periodic ($MN$ is the frame length) after CP removal, such that DZT can be employed for receiver processing, yielding an effective DD domain channel matrix with block diagonal structure~\cite{Raviteja2019practical}. There are some interesting variations of reduced-CP OTFS. For example, it is reported in~\cite{Raviteja2018interference,pandey2021low} that padding zeros instead of adding a CP results in a more structured effective DD domain channel matrix, at the cost of a small power loss.

\subsection{Window Design for SFFT-based OTFS}
An appealing advantage of SFFT-based OTFS is that it facilitates TF domain window design~\cite{wei2021transmitter}, which introduces additional DoFs for further improvements of the channel estimation and data detection performance compared to the commonly used rectangular window.
The windowing at the transmitter can be interpreted as power allocation in the TF domain, while the windowing at the receiver causes colored noise\cite{wei2021transmitter}.
If channel state information (CSI) is available at both transmitter and receiver, the transmitter window can be optimized for minimization of the detection mean squared error (MSE). The obtained solution can be interpreted as a mercury/water-filling power allocation, where the mercury is filled first, before water is poured to pre-equalize the doubly selective TF domain channel\cite{wei2021transmitter}.
If CSI is not available at the transmitter, fixed window designs, such as the Dolph-Chebyshev (DC) window\cite{wei2021transmitter}, in the TF domain can enhance channel sparsity and thus improve channel estimation performance, enabling a smaller guard space overhead.
We refer interested readers to \cite{wei2021transmitter} for a more detailed discussion of window designs for OTFS modulation.

\subsection{Pulse Shaping for DZT-based OTFS}
Different from SFFT-based OTFS, DZT-based OTFS directly converts the DD domain signal into the time-delay (TD) domain without converting the signal first into the TF domain. DZT-based OTFS transmitters generally comprise an IDZT module and a pulse shaping filter ${g_{{\rm{tx}}}}\left( t \right)$. According to (15) in Part I and~\cite{lampel2021orthogonal},
the discrete DD domain equivalent transmitted symbols can be obtained via the DZT of the samples of the TD domain transmit signal $s\left( t \right)$,
such that
\begin{align}
{\cal D}{{\cal Z}_{{s}}}\left[ {l,k} \right] = \sqrt {MN} {X_{{\rm{DD}}}}\left[ {l,k} \right]{\cal D}{{\cal Z}_{g_{\rm tx}}}\left[ {l,k} \right],\label{DD_equivalent_with_pulse}
\end{align}
where information symbol ${X_{{\rm{DD}}}}\left[ {l,k} \right]$ is the $(l,k)$-th element of the DD domain information matrix ${\bf X}_{\rm DD}$ of size $M \times N$, with $l \in \left\{ {0,...,M - 1} \right\}$ and $k \in \left\{ {0,...,N - 1} \right\}$. In~\eqref{DD_equivalent_with_pulse}, ${{\cal DZ}_{g_{\rm tx}}}$ denotes the DZT of
vector ${\bf g}_{\rm tx}$ containing the \emph{periodically extended} pulse samples, i.e., for the $k$-th element of ${\bf g}_{\rm tx}$, we have $g_{\rm tx}\left[ k \right] \buildrel \Delta \over = {g_{{\rm{tx}}}}\left( {\frac{{{{\left[ k \right]}_{MN}}}}{M}T} \right)$, $k \in {\mathbb Z}$, where ${\left[ \cdot \right]}_{N}$ denotes the modulo operation with respect to (w.r.t.) $N$.

\begin{figure}
\centering
\includegraphics[width=3.5in]{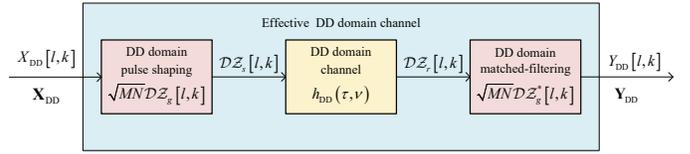}
\caption{Block diagram of the equivalent model for DZT-based OTFS.}\vspace{-5mm}
\label{Zak_equivalent_model}
\centering
\end{figure}

The literature on OTFS pulse shape design is not mature yet, however, we may still provide some intuition for pulse shape design in the DD domain.
In fact,~\eqref{DD_equivalent_with_pulse} suggests an interesting interpretation of pulse shape design in the DD domain, where the pulse shaping can be viewed as a point-wise multiplication.
Notice that the DZT is defined for $MN$-periodic sequences. With the reduced-CP scheme mentioned in the previous subsection, the overall DZT-based OTFS communication system can be equivalently modelled as in Fig.~\ref{Zak_equivalent_model}.
In Fig.~\ref{Zak_equivalent_model}, we assume that the same pulse is employed for
both transmit pulse shaping and receive matched-filtering, and its \emph{periodically extended} sample vector is given by $\bf g$, i.e., ${{\bf g}_{{\rm{tx}}}}={{\bf g}_{{\rm{rx}}}}={{\bf g}}$.
Thus, the pulse shape design for OTFS may be formulated as an optimization problem that aims to optimize the effective DD domain channel ${{\cal DZ}_g}\left[ {l,k} \right]h_{\rm DD}\left[ {l,k } \right]{\cal DZ}_g^*\left[ {l,k } \right]$, $l \in \left\{ {0,...,M - 1} \right\}$ and $k\in \left\{ {0,...,N - 1} \right\}$, where $h_{\rm DD}\left[ {l,k } \right]$ denotes the samples of the continuous DD domain channel response $h_{\rm DD}\left( {\tau ,\nu } \right)$.

An effective tool used for pulse shape design is the \emph{cross ambiguity function}. The ambiguity function characterizes the correlation between two time domain signals w.r.t. delay variable $\tau$ and Doppler variable $\nu$, and is defined as follows~\cite{Raviteja2018interference}
\begin{align}
{A_{x,y}}\left( {\tau ,\nu } \right) \buildrel \Delta \over = \int_{ - \infty }^\infty  {x\left( t \right)} {y^*}\left( {t - \tau } \right){{\rm{e}}^{ - j2\pi \nu \left( {t - \tau } \right)}}{\rm{d}}t. \label{AF}
\end{align}
In the literature, a pulse is referred to as an \emph{ideal} pulse, if it satisfies the bi-orthogonality condition~\cite{Hadani2017orthogonal}, i.e.,
\begin{align}
{A_{{g_{\rm tx}},g_{\rm rx}}}\left( {nT,\frac{m}{T}} \right) = \delta \left[ n \right]\delta \left[ m \right],
\end{align}
where $\delta \left[ {\cdot} \right]$ is the Dirac delta function.
Note that the ideal pulse is defined on a TF domain grid, which implies two-dimensional (2D) orthogonality between TF domain transmitted symbols. However, a pulse satisfying the bi-orthogonality condition in the TF domain may not have ideal properties in the DD domain, where the grid (corresponding to the DD resolution) is defined differently.
For the design of the pulse shape in the DD domain, we may exploit the relation between a product of DZTs and the ambiguity function. In particular, as shown in~\cite{Bolcskei1994Gabor}, the product of two DZTs can be expanded into a 2D Fourier series w.r.t. the sampled cross ambiguity function, which could be exploited for pulse shape design.
More specifically, ${{\cal DZ}_g}\left[ {l,k} \right]h_{\rm DD}\left[ {l,k } \right]{\cal DZ}_g^*\left[ {l,k } \right]$ can be optimized by leveraging the cross ambiguity function with the objective to promote certain properties, such as improved channel sparsity and larger Euclidean distance.



\section{Receiver Design}
In this section, we consider the receiver design for OTFS systems. Due to the space limitation, we focus on DZT-based OTFS.

\subsection{Channel Estimation}
Different from its OFDM counterpart, OTFS channel estimation is usually performed in the DD domain rather than the TF domain as this allows the exploitation of
the appealing properties of DD domain channel responses, such as sparsity, compactness, separability, and quasi-static behaviour~\cite{Zhiqiang_magzine}.
A commonly used channel estimation approach for OTFS may be the one published in~\cite{Raviteja2019embedded}, which only requires one embedded pilot symbol in the DD domain.
Specifically, a sufficiently large guard interval is applied around the pilot to facilitate the acquisition of the delay and Doppler responses. As the DD domain relationship between the transmitted signal and the channel response corresponds to a 2D circular convolution as discussed in Part I, the embedded pilot is smeared over several DD grid points around the original location.
Therefore, the channel can be estimated by simply checking the received signal's values around the DD grid point where the pilot was embedded.

Channel estimation based on compressed sensing methods has also been considered for OTFS systems. Compressed sensing is suitable for sparse signal recovery, where the number of measurements is much smaller than the number of unknown parameters. Therefore, compressed sensing-based channel estimation is well-suited for OTFS with fractional delays or/and fractional Doppler shifts~\cite{Wei2022off}.
For instance, the
authors in~\cite{Wenqian2019channel} proposed a three-dimensional (3D) structured orthogonal matching pursuit (OMP) algorithm to estimate the delay-Doppler-angle domain channel by exploiting the underlying 3D structured sparsity.
A 3D Newtonized OMP (NOMP) algorithm was proposed in~\cite{Muye2021new_path}, which exploits the fractional components in the Doppler and angle domains via Newton's method.
Furthermore, channel estimation based on sparse Bayesian learning (SBL) techniques has been recently proposed~\cite{Wei2022off}, and was shown to achieve a better error performance compared to OMP-based schemes.

\begin{figure}
\centering
\includegraphics[width=2.8in]{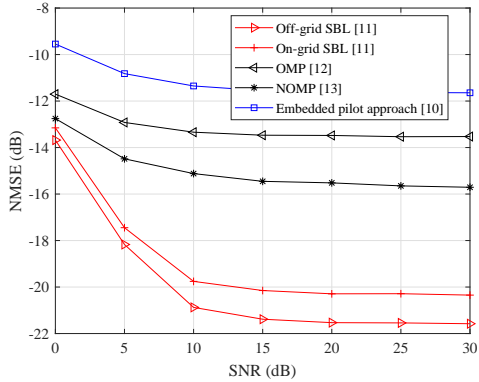}\vspace{-2mm}
\caption{Channel estimation performance comparison for SBL~\cite{Wei2022off}, OMP~\cite{Wenqian2019channel}, NOMP~\cite{Muye2021new_path}, and the conventional embedded pilot approach~\cite{Raviteja2019embedded}.}\vspace{-5mm}
\label{CS_performance}
\centering
\end{figure}

We present a performance comparison of the above-mentioned channel estimation schemes w.r.t. the signal-to-noise ratio (SNR) in Fig.~\ref{CS_performance}, where $M=N=32$, $P=5$, $l_{\rm max}=4$, and $k_{\rm max}=3$, respectively, and the fading coefficients are generated according to the exponential power delay profile with exponent $0.1$. As can be observed, both on-grid SBL and off-grid SBL~\cite{Wei2022off} outperform OMP~\cite{Wenqian2019channel}, NOMP~\cite{Muye2021new_path}, and the conventional embedded pilot approach~\cite{Raviteja2019embedded} in terms of the normalized mean squared error (NMSE). Furthermore, off-grid SBL achieves roughly $1$ dB NMSE gain over on-grid SBL in the high SNR regime because off-grid SBL can model the effects of fractional Doppler components. Meanwhile, we also observe that NOMP~\cite{Muye2021new_path} achieves roughly $2$ dB NMSE gain over OMP~\cite{Wenqian2019channel}, since NOMP can refine the delay and Doppler shift estimation via Newton's method.

\subsection{Signal Detection}

For OTFS, conventional detectors can be used, such as the minimum mean square error (MMSE) detector. Thanks to the properties of the effective DD domain channel matrix, MMSE detection can be implemented with linear time complexity~\cite{Chockalingam2020low_comp}.

Apart from MMSE detection, MPA~\cite{Raviteja2018interference,li2021hybrid,Yuan2020simple} has also been widely applied for OTFS detection.
Let us briefly introduce MPA from an MAP detection point of view~\cite{li2021hybrid}.
In the case of integer delay and Doppler shifts, assuming an ideal pulse, the $(l,k)$-th element $Y_{\rm DD}\left[ {l,k} \right]$ of the DD domain received symbol matrix ${\bf Y}_{\rm DD}$, for $l \in \left\{ {0,...,M - 1} \right\}$ and $k\in \left\{ {0,...,N - 1} \right\}$, is given as follows~\cite{Raviteja2018interference}
\begin{align}
Y_{\rm DD}\left[ {l,k} \right] &\!=\! \sum\nolimits_{i = 1}^P {{h_i}{e^{ - j2\pi {\nu _i}{\tau _i}}}X_{\rm DD}\!\!\left[ {{{\left[ {l - l_i} \right]}_M},{{\left[ {k - k_i} \right]}_N}} \right]}\notag\\
&  + Z_{\rm DD}\left[ {l,k} \right], \label{IO_ideal_integer}
\end{align}
where $P$ is the number of resolvable paths, $h_i$, ${\tau _i}$, and ${\nu _i}$ are the fading coefficient, the delay, and the Doppler shift associated with the $i$-th path, respectively, while ${l _i}$ and ${k _i}$ are the corresponding delay and Doppler indices, as defined in Part I.
For ease of presentation, let us define the following sets
\begin{align}
\mathbb{H}^{\left( i \right)}&\buildrel \Delta \over = \left\{ {{h_j}\left| {1 \le j \le P,j \ne i} \right.} \right\},\notag\\
\mathbb{Y}_{l,k} &\buildrel \Delta \over =  \left\{ {Y_{\rm DD}\left[{{\left[ {l + l_i} \right]}_M},{{\left[ {k + k_i} \right]}_N}\right]\big| {1 \le i \le P} } \right\}, \text{and}\notag\\
\mathbb{X}_{l,k}^{\left( i \right)}\!&\buildrel \Delta \over = \!\!
 \left\{ {{X_{{\rm{DD}}}}\!\!\left[ {\left[ {l\! +\! {l_i} \!- \!{l_j}} \right]_M\!,\!\left[ {k \!+\! {k_i}\! - \!{k_j}} \right]_N} \right]\!\left| {1 \le j \le P,j \ne i} \right.} \right\}\notag,
\end{align}
where the $j$-th element of $\mathbb{H}^{\left( i \right)}$, $\mathbb{Y}_{l,k}$, and $\mathbb{X}_{l,k}^{\left( i \right)}$, are denoted by $\mathbb{H}^{\left( i \right)}[j]$, $\mathbb{Y}_{l,k}[j]$, and $\mathbb{X}_{l,k}^{\left( i \right)}[j]$, respectively.
According to~\eqref{IO_ideal_integer}, it can be shown that set $\mathbb{Y}_{l,k}$ contains $P$ received symbols that are associated with DD domain transmitted symbol $X_{\rm DD}\left[ {l,k} \right]$,
while set $\mathbb{X}_{l,k}^{\left( i \right)}$ contains $P-1$ DD domain transmitted symbols that are related to received symbol $\mathbb{Y}_{l,k}\left[ i \right]$, i.e., ${Y_{\rm DD}\left[{{\left[ {l + l_i} \right]}_M},{{\left[ {k + k_i} \right]}_N}\right]}$.
In fact, the \emph{a posteriori} probability $\Pr \left\{ {{X_{{\rm{DD}}}}\left[ {l,k} \right]|{{\bf{Y}}_{{\rm{DD}}}}} \right\}$ can be factorized based on a graphical model, where the nodes and calculations can be characterized by $\mathbb{H}^{\left( i \right)}$, $\mathbb{Y}_{l,k}$, and $\mathbb{X}_{l,k}^{\left( i \right)}$, respectively, as shown in~\cite{li2021hybrid}. Due to the page limitation, we cannot provide the implementation details for MPA. However, we note that the main idea of MPA is to pass messages among the connected nodes iteratively in a graphical model, such that the target probability, e.g., the \emph{a posteriori} probability, is approximately calculated after a sufficient number of iterations.

\begin{figure}
\centering\vspace{-3mm}
\includegraphics[width=2.8in]{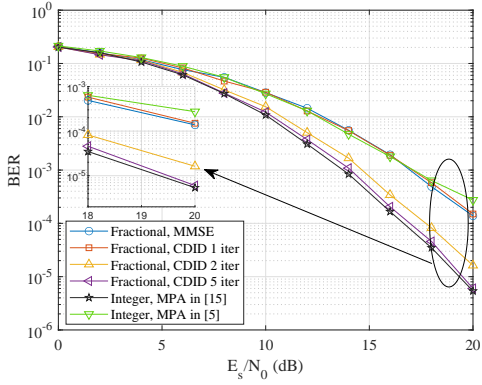}\vspace{-3mm}
\caption{BER comparison of MMSE detection, MPA~\cite{li2021hybrid}, MPA~\cite{Raviteja2018interference}, and CDID.}\vspace{-5mm}
\label{Detection_P4}
\centering
\end{figure}

Note that the MPA designed based on~\eqref{IO_ideal_integer} assumes integer delays and Doppler shifts. In the fractional Doppler case, cross domain iterative detection (CDID) proposed in~\cite{li2021cross} has been shown to achieve a near-optimal performance with reduced complexity{\footnote{We note that detection algorithms for OTFS are generally designed specifically with different channel conditions in mind. For example, the MPA algorithm reported in~\cite{li2021hybrid} is not suitable for fractional Doppler shifts, as its detection complexity would become prohibitively high in this case.}}. CDID employs simple estimation/detection schemes in both the TD and DD domains and iteratively updates the extrinsic information via the unitary transformations between the TD and DD domains. Fig.~\ref{Detection_P4} shows the bit error ratio (BER) performance of OTFS transmission for conventional MMSE detection, MPA in~\cite{li2021hybrid}, MPA in~\cite{Raviteja2018interference}, and CDID, where we adopted $M=32$, $N=16$, $P=4$ and the fading coefficients are generated based on a uniform power delay profile with $l_{\rm max}=10$ and $k_{\rm max}=5$. The MPA in~\cite{Raviteja2018interference} and conventional MMSE detection achieve roughly the same BER, which also coincides with that of the first iteration of CDID. Furthermore, as the number of iterations increases, CDID gradually approaches the performance of the MPA in~\cite{li2021hybrid} with integer Doppler shifts, which is approximately the MAP detection performance. This observation suggests that CDID bridges the performance gap between MMSE and MAP as the number of iterations increase, which indicates that CDID achieves a favorable performance-complexity tradeoff. For more details regarding the performance analysis of CDID, we refer to~\cite{li2021cross}.

\vspace{-2mm}
\section{Performance Analysis of OTFS Modulation}
In this section, we analyze the performance of OTFS modulation and draw comparisons with OFDM. 
To this end, we consider system representative parameters to facilitate our discussion. In practice, all parameters have to be selected carefully according to the underlying channel conditions, of course.

\vspace{-2mm}
\subsection{Diversity Gain vs. Coding Gain}
The diversity gain characterizes the exponential scaling of the error performance w.r.t. the SNR in the high SNR regime.
OFDM requires channel coding to extract the diversity gain offered by multipath channels. In contrast, OTFS has the potential to exploit the full channel diversity without channel coding~\cite{Surabhi2019on,Raviteja2019effective,li2020performance}.
Nevertheless, channel coding will further improve the error performance of OTFS.
In particular, it is shown in~\cite{li2020performance} that the unconditional pair-wise error probability (PEP) of coded OTFS modulation over Rayleigh fading channels can be approximately upper-bounded by
\begin{equation}
\Pr\left( { {{\bf{x}},{\bf{x'}}} } \right)\mathbin{\lower.3ex\hbox{$\buildrel<\over
{\smash{\scriptstyle\sim}\vphantom{_x}}$}}{\left( {\frac{{d_{\rm{E}}^2\left( {\bf{e}} \right)}}{P}} \right)^{ - P}}{\left( {\frac{{{E_s}}}{{4{N_0}}}} \right)^{ - P}},\label{diversity_coding_tradeoff}
\end{equation}
where $\Pr\left( { {{\bf{x}},{\bf{x'}}} } \right)$ denotes the probability that DD domain transmitted sequence $\bf x$ is mistakenly detected as $\bf x'$, and the Euclidean distance between $\bf x$ and $\bf x'$ is ${d_{\rm{E}}^2\left( {\bf{e}} \right)}$. In~\eqref{diversity_coding_tradeoff}, the SNR exponent, i.e., the diversity gain, is equal to the number of resolvable paths of the underlying wireless channel $P$. On the other hand, the term ${{d_{\rm{E}}^2\left( {\bf{e}} \right)} \mathord{\left/
 {\vphantom {{d_{\rm{E}}^2\left( {\bf{e}} \right)} P}} \right.
 \kern-\nulldelimiterspace} P}$ is referred to as the coding gain, indicating the SNR gain achieved with channel coding~\cite{li2020performance}.
Two interesting observations can be obtained from~\eqref{diversity_coding_tradeoff}. Firstly, the PEP upper-bound does not depend on the delays and Doppler shifts, which implies that OTFS modulation causes ``channel hardening''. This is because each DD domain transmitted symbol experiences the fluctuation of the entire TF domain channel response thanks to the employed ISFFT.
Secondly, there is a tradeoff between diversity gain and coding gain for OTFS. In particular,~\eqref{diversity_coding_tradeoff} indicates that the diversity gain of OTFS improves with the number of resolvable paths $P$, while the coding gain declines. This observation suggests a rule-of-thumb for code design, i.e., the
Euclidean distance between transmitted sequences should be maximized, which actually aligns with the code design criterion for the additive white Gaussian noise (AWGN) channel as a consequence of the ``channel hardening'' effect.
\begin{figure}
\centering
\includegraphics[width=2.8in]{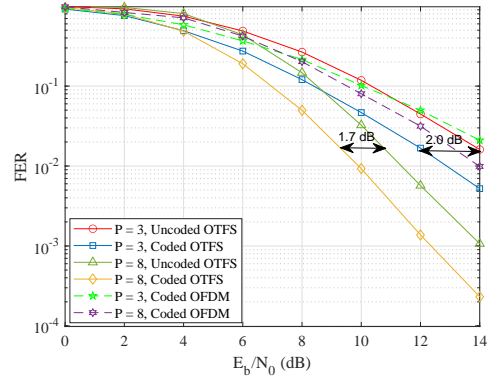}\vspace{-3mm}
\caption{Comparison of FER performances of coded and uncoded OTFS and OFDM.}\vspace{-5mm}
\label{div_coding_tradeoff}
\centering
\end{figure}
Fig.~\ref{div_coding_tradeoff} depicts the frame error rate (FER) performances of coded and uncoded OTFS and OFDM modulation with maximum-likelihood (ML) detection, where we apply the half-rate (3,1) feedforward convolutional code and binary phase shift keying (BPSK). From the figure, we observe that for a larger number of resolvable paths, the coding gain for OTFS decreases, e.g., from $2.0$ dB to $1.7$ dB, while the diversity gain increases, which is consistent with our discussions based on~\eqref{diversity_coding_tradeoff}. Furthermore, we also notice that the diversity gain of coded OTFS with $P = 8$ is larger than that of coded OFDM, which suggests that coded OTFS is a more attractive option for reliable communication over multipath fading channels than coded OFDM.

\vspace{-3mm}
\subsection{Achievable Rate Performance}
\begin{figure}
\centering
\includegraphics[width=2.8in]{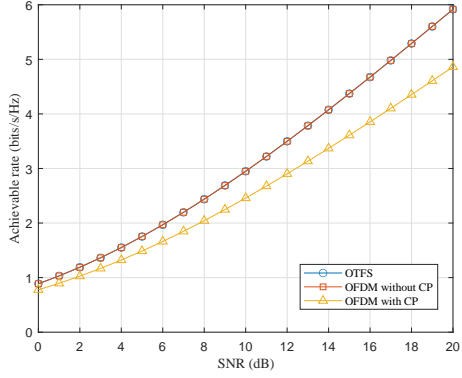}\vspace{-3mm}
\caption{Comparison of the achievable rates of OTFS, OFDM without CP, and OFDM with CP, where perfect CSI is assumed and $P=4$.}\vspace{-4mm}
\label{Achievable_rate}
\centering
\end{figure}

The achievable rate is an important performance metric characterizing how much information can be reliably transmitted over a channel with given resources. The achievable rates of OTFS and OFDM have been compared in~\cite{RezazadehReyhani2018analysis,Chong2022achievable}. 
We present the achievable rate performance for both OTFS and OFDM in Fig.~\ref{Achievable_rate}, where we assume that perfect CSI is available at the receiver side and the achievable rate is calculated based on
\begin{align}
\setcounter{equation}{5}
R = \frac{1}{{MN}}{\log _2}\det \left( {{{\bf{I}}_{MN}} + {\rm{SNR}}{{\bf{H}}^{\rm{H}}}{\bf{H}}} \right).\label{achievable_rate_formula}
\end{align}
In~\eqref{achievable_rate_formula}, ${{\bf{I}}_{MN}}$ denotes the identity matrix of size $MN$, ${\rm{SNR}}$ denotes the operating SNR, ${\rm det}(\cdot)$ denotes the determinant, $(\cdot)^{\rm H}$ denotes the Hermitian conjugate, and ${\bf{H}}$ stands for the \emph{effective channel matrix} for reduced-CP OTFS or OFDM with and without CP (the CP length equals $l_{\rm max}$), as given in~\cite{Raviteja2019practical} and~\cite{RezazadehReyhani2018analysis}, respectively.
We set $M=32$ and $N=16$, and assume $P=4$ independent resolvable paths with maximum delay and Doppler indices given by $l_{\rm max}=5$ and $k_{\rm max}=5$, respectively.
As can be observed from Fig.~\ref{Achievable_rate}, reduced-CP OTFS and OFDM provide almost the same achievable rate, while OFDM with CP clearly suffers from a rate degradation due to the CP insertion. The intuition behind this observation is that the transformation between the TF domain and the DD domain is unitary, and thus, does not affect matrix determinants, leading to the same achievable rate.
However, the unitary property of the domain transformation may not hold in the multiuser case, where only a limited number of resource blocks can be allocated to each user. In this case, it has been shown that OTFS yields an overall achievable rate gain compared to OFDM if practical successive interference cancelation (SIC) detection is employed at the receiver~\cite{Chong2022achievable}.

An alternative, more practical performance metric is the pragmatic capacity, defined as the achievable rate of the channel
induced by the signal constellation and the soft-output of the detector~\cite{Lorenzo2021otfs}. The authors in~\cite{Lorenzo2021otfs} showed that OTFS transmission enjoys a better pragmatic capacity performance compared to OFDM over static channels with practical channel estimation and detection schemes, thanks to the smaller signaling overhead. Furthermore, the pragmatic capacity of OFDM is
very sensitive to the Doppler effect~\cite{Lorenzo2021otfs}, such that OTFS has a clear advantage in high-mobility channels.


\vspace{-2mm}
\section{Conclusions and Future Research Directions}
In this letter, we reviewed OTFS transceiver design principles, including CP insertion, pulse shaping, channel estimation, and signal detection. We also discussed the diversity gain and achievable rate of OTFS systems.
It is worth pointing out that OTFS transceiver design still faces many practical issues. For example, OTFS receivers may induce a long latency as the demodulation can only be carried out once the whole block of TF symbols is received due to the symbol spreading from the DD domain to the TF domain. Furthermore, without a carefully designed pulse shape, OTFS may cause high out-of-band emissions and other practical issues. Therefore, low latency receiver and pulse designs are important research topics for facilitating practical OTFS implementation.

\vspace{-3mm}
\bibliographystyle{IEEEtran}
\bibliography{OTFS_references}

\end{document}